# First observation of a reactor-status effect on the $\beta^+$ decay rate of $^{22}$Na

Robert de Meijer,[1,2,*] Albert Zondervan,[3,†] Jan Stegenga,[4] Steph Steyn,[5] Robbie Lindsay,[2] Milton van Rooy,[6] Marco Tijs,[7] and Han Limburg[7]

[1]*Stichting EARTH, Weehorsterweg 2, 9321 XS Peize, Netherlands*
[2]*Department of Physics and Astronomy, University of the Western Cape, Bellville, South Africa*
[3]*GNS Science, PO Box 31-312, Lower Hutt 5040, New Zealand*
[4]*INCAS³, Dr. Nassaulaan 9, 9401 HJ Assen, Netherlands*
[5]*Eskom Holdings SOC Ltd, Kernkrag, South Africa*
[6]*National Metrology Institute of South Africa, Cape Town, South Africa*
[7]*MEDUSA Sensing BV, 9723 JV Groningen, Netherlands*

In the search for an electron antineutrino detection method with sensitivity below the 1.8 MeV threshold for the inverse beta decay reaction, β-decay counting experiments with ca. 3 kBq $^{22}$Na and $^{60}$Co sources were conducted at unit #1 (2.775 GW$_{th}$) of the Koeberg Nuclear Power Station in South Africa. The goal was to determine if the rate of decay is measurably influenced by a change between the ON–OFF status of such a reactor. The experimental setup consisted of a single NaI crystal to measure de-excitation and annihilation photons associated with β-decay. Its volume and well shape were purposely chosen to use coincidence summing in the interval 170–2452 keV to differentiate between electron capture and $\beta^+$ emission in $^{22}$Na. The Pb-shielded setup was placed in the seismic vault underneath the containment building, thereby shielded from the reactor core by 8 m of uninterrupted concrete. Background radiation, responsible for ca. 1% of the total countrate with either source placed in the NaI well, increased by merely 3% when the reactor status changed from OFF to ON. This small increase is semi-quantitatively explained by fast neutrons exciting $^{208}$Pb nuclei throughout the Pb castle. Offline analysis of measured pulseheight spectra comprised background subtraction, correction for natural decay, grouping into daily averages, energy calibration, and integration over three energy regions-of-interest (TOT, MED, HI). Subsequently, normalized countrates were parameterized to jointly describe the time dependence of two instrumental effects and a reactor-status step function in a least-squares regression analysis. For $^{22}$Na two measurement series were made, each covering an ON–OFF–ON cycle of the reactor. The following fractional countrate (activity) changes in the step from reactor OFF to ON were obtained: $(\Delta A/A)_{TOT} = [-3.02 \pm 0.14(stat) \pm 0.07(syst)] \times 10^{-4}$, $(\Delta A/A)_{MED} = [+1.44 \pm 0.42(stat) \pm 0.07(syst)] \times 10^{-4}$, and $(\Delta A/A)_{HI} = [-2.70 \pm 0.26(stat) \pm 0.04(syst)] \times 10^{-4}$. The systematic errors are governed by uncertainty in the difference between background spectra during reactor ON and OFF. However, the uncertainty budget is incomplete because it does not contain the possible influence from environmental factors and the finite stability of the MCA clock-oscillator. No reactor status dependence was observed with the $^{60}$Co source in the counter. The corresponding cross sections with their statistical uncertainty are $(1.55 \pm 0.07) \times 10^{-25}$ cm$^2$ for EC+$\beta^+$ decay in $^{22}$Na and $(0.5 \pm 1.5) \times 10^{-26}$ cm$^2$ for $\beta^-$ decay of $^{60}$Co. The negative sign for TOT and HI activity changes in $^{22}$Na points to an antineutrino related interference effect on the $\beta^+$ decay of $^{22}$Na and rules out reactor neutron induced reactions.

## I. INTRODUCTION

Efficient detection of low-energy electron antineutrinos is of great relevance to both fundamental and applied physics. With nuclear power reactors being an intense and flexible source of that type of neutrinos, several open questions in neutrino physics can in principle be addressed, such as the existence of sterile neutrinos, short-baseline flavor oscillations, and the reactor antineutrino anomaly [1-5]. The development of new methods to monitor reactor fuel composition and burn-up are key to the efforts by the IAEA in its mission to implement reactor safeguards and promote anti-proliferation measures [6-9].

All past and current experiments to study electron antineutrinos from fission reactors utilize the Inverse Beta Decay (IBD) reaction on the free proton. The kinematics of this reaction is well known [10]. Due to the negative Q-value for this reaction, detection via this method is limited to neutrino energies above the threshold energy: $E_{\bar{\nu}} > -Q = 1.8 \, MeV$. Good discrimination against background events is achieved by requiring the coincident detection of the prompt signal from the emitted positron plus annihilation quanta and the delayed signal from capture of the emitted neutron [11]. In principle, this IBD technique lends itself to full reconstruction of the interaction vertex and the momentum vector of the incoming antineutrino.

---

[*] rmeijer@geoneutrino.nl
[†] a.zondervan@gns.cri.nz

Despite the advance of knowledge made through use of the IBD reaction, widespread deployment of detectors based on it remains a formidable challenge because useful reactor neutrino countrates demand detector volumes of at least a few cubic meters as the IBD cross section is on the order of $10^{-42}$ cm$^2$. Coherent antineutrino–nucleus elastic scattering has a significantly larger cross section [12], but observation has remained elusive due to the difficulty in detecting the ≈1 keV nuclear recoils [13]. Capture on beta-decaying nuclei has been suggested as a mechanism to gain access to detecting low-energy antineutrinos [14], but also this approach has so far not been realized.

Our approach towards detecting antineutrinos was triggered by the hypothesis that the $\beta^-$ decay rates of $^{32}$Si and $^{226}$Ra are influenced by solar neutrinos, as possible explanation of observations that the variations in these rates, after correcting for decay, have an inverse relationship with the Sun–Earth distance [15]. Instead of attempting to detect neutrinos individually through any of the abovementioned reaction mechanisms, we aim at detecting antineutrinos indirectly, i.e. through their influence on β-decay rates. In our earlier attempts with decay-counting a $^{22}$Na source, using HPGe and LaBr$_3$ detectors, we failed to observe a non-zero reactor-status effect [16-18]. Recently, Barnes et al. have reported on an experiment to search for reactor-status effects in the decay rates of $^{54}$Mn, $^{22}$Na, and $^{60}$Co, using the High Flux Isotope Reactor (HFIR) at Oak Ridge National Laboratory [19]. Partly due to the high level of background radiation [20], the sensitivity of their setup was insufficient to exclude perturbations less than one or two parts in $10^4$.

In this investigation we continue this search with an improved setup using a well-type NaI detector and a considerably larger antineutrino flux (Sec. II), and in a lower background radiation environment (Sec. III). Analysis of the Dec-2012–Dec-2015 measurements on three nuclides are presented in Secs. IV–VI, followed by a discussion on the robustness and possible interpretation of the non-zero reactor-status effect in Sec. VII. We draw our conclusions in Sec. VIII.

## II. METHOD AND INSTRUMENTAL FACILITIES

Low-energy electron antineutrinos are a product of the decay of neutron-rich nuclides produced in actinide fission and spallation reactions. To study how strongly that class of neutrino interacts with proton-rich β-decaying nuclides, characteristic γ-rays associated with their decay can be utilized. With a nuclear power reactor to provide the antineutrinos, a γ-ray counting setup placed nearby can monitor the decay of a β-radioactive source during extended periods of different stages of reactor operation. Countrate determinations then allow the effect of reactor-status changes on decay rates to be quantified. The use of specific intervals in the γ-ray energy spectrum gives the possibility to distinguish between decay branches. In this section, we describe how these ideas were combined in an experimental setup and the design of the offline analysis of measurement data.

### A. Reactor facility

The measurements were carried out at unit #1 of the Koeberg Nuclear Power Station (Eskom), about 30 km north of Cape Town, South Africa. Koeberg Station has two pressurized water reactor (PWR) units, each of nominal 970 MW$_e$ (gross) and fueled with low-enriched uranium. Refueling and maintenance are scheduled on an 18-month cycle. Fissile material is distributed over the fuel rods in such a way as to optimize heat production and fuel burn-up.

Each containment building has a ca. 1 m thick concrete roof and a ca. 8 m thick concrete floor. That floor rests on a neoprene layer on top of 2 x 2 x 2 m$^3$ concrete blocks which, in turn, rest on a 2 m thick concrete floor. The area between the two floors is called the seismic vault. A schematic cross section of the relevant part of the containment building and the seismic vault underneath is shown in Fig. 1. The detector in our measurements is located in the seismic fault, next to the concrete block right under the reactor vessel. The distance to the center of the reactor core is ca. 15 m. There are no feedthroughs in the containment floor. The seismic vault is not part of the radiation controlled zone and hand-held monitors did not indicate enhanced radiation levels there. The seismic vault has openings to the outside air for natural ventilation and the air temperature at the location of the setup does not vary by more than 2°C throughout the year.

### B. Antineutrino flux

At present we do not consider the shape of the antineutrino energy distribution nor how it changes during fuel burn-up: We limit ourselves to estimating the total flux and the flux (density) at the detector. Koeberg's fuel composition at the start and end of the cycle is given in Table I in terms of fission neutron and fissioning nuclide fractions. The contributions from [235]U and [239]Pu change significantly during burn-up. Using the known average number of antineutrinos produced per fission [21], it is calculated for the composition as given in Table I that the total antineutrino flux drops by only 2% during the entire cycle.

The flux is estimated by assuming that all production takes place at the center of the reactor core cylinder at 15.2 ± 0.5 m from the detector, by assuming 202 MeV energy release per fission, a 2.775 ± 0.010 GW$_{th}$ heat generation, and by using the average number of antineutrinos produced per fission as shown in Table I. This leads to an antineutrino flux from unit #1 at the center of the detector equal to 1.65 x 10$^{13}$ cm$^{-2}$ s$^{-1}$. Our gross simplification of the actual fuel distribution inside the core renders this estimate uncertain by 30%.

The thrust of using a nuclear fission reactor as a source of antineutrinos lies in the ability to quickly ramp the flux up and down. Our first goal is to establish if the activity of a source of a β-decaying nuclide responds to reactor state changes. For that purpose, uncertainty in the antineutrino flux is largely irrelevant. Of course, accuracy of the flux calculation will be important when interpreting changes in decay rates, if found, in terms of a new or known physics phenomenon.

### C. Detector setup

The detector is a well-geometry NaI crystal with outer cylindrical dimensions $\phi = l = 10.2\,cm$ and $\phi = 1.2\,cm$, $l = 5.1\,cm$ for the coaxial well. A quartz light guide couples the crystal to a 7.4 cm diameter, low-K window photomultiplier tube (PMT). The detector assembly, made by Scionix, has configuration 102-BP-102/3-M-X. The detector is surrounded by a Pb castle at least 7.5 cm thick and lined on the inside by a 1 mm thick Cu layer. The top part of the castle consists of a dove-tailed sliding door. At the bottom of the castle, the anode end of the PMT protrudes through a $\phi = 8.5\,cm$ hole, leaving some space for natural ventilation. Shielding from upwardly directed background radiation is provided by two Pb discs 5 cm thick and 25 cm in diameter. The structure is supported by a steel frame that stands on the floor of the seismic vault.

The [22]Na and [60]Co sources are mounted on the end of polycarbonate rods to achieve sufficient repeatability in the source–crystal geometry during multiple exchanges of sources. To protect the crystal from positrons, electrons, and X-rays emitted by the source, each rod with source is wrapped in 1 mm thick Al foil and surrounded by ca. 3 mm thick Pb foil. The source is placed in the center of the crystal. At the start of the measurements the source strengths are 2 and 3 kBq for [22]Na and [60]Co, respectively. The [50]V source consists of a $\phi = 11\,mm$, $l = 50\,mm$ rod of natural vanadium, not wrapped in foils. From the low natural abundance and the long half-life of [50]V, the only naturally occurring radioactive vanadium isotope, it can be calculated that the strength of the [50]V source is 0.13 mBq.

The multi-channel analyzer (MCA) is an integral part of the base electronics unit of the PMT. This unit is connected to a laptop computer (Lenovo) for power and data transfer. For the [22]Na measurements a Venus digital MCA (Itech Instruments) was used, while for the [60]Co and [50]V measurements a Scintispec analog MCA (FLIR Systems) was used. Both systems are equipped with on-line gain stabilization.

A consequence of measuring the countrate of a radioactive source is that its strength reduces over time and thus that the average time separation between the pulses increases. The effects that pulse pile-up and baseline shift have on the energy spectrum therefore slowly diminish over time. Peaks in the spectrum tend to shift to lower channels with time due to the diminishing average height of the preceding pulse. Since this particular effect is energy independent, it appears as a gradual shift of the whole spectrum towards lower pulse height. Online gain stabilization is well suited to compensate for temperature changes affecting photoelectron multiplication in the PMT. However, maintaining one preselected peak at a fixed channel number introduces non-linearity when attempting to compensate for changes in pulse pile-up and baseline shift across the whole spectrum. See the thorough discussion of this unwanted effect in [17,18].

### D. Coincidence summing

In view of our aim to measure a possible change in countrate with high precision and accuracy, we opted for the well counter geometry because of its solid angle, a reasonable energy resolution and high detection efficiency. By setting wide Regions-of-Interest (RoIs) we enhance Poisson counting precision and avoid systematic uncertainties due to making assumptions about the shape of continua under peaks. It is clear that in this experiment energy resolution is less important than stability of the energy scale. The ability to detect small countrate variations over long periods depends largely on accurate calibration of each pulse-height spectrum.

Placing the source near the bottom of the well in the crystal not only boosts the detection solid angle but also enhances the probability of summing the energy depositions of multiple γ-rays associated with the decay of a single nucleus: the so-called coincidence summing. The moderate time-resolution of the NaI detector allows prompt cascade and 511 keV γ-rays following $β^+$ decay, assuming they interacted within the crystal, to be registered as a single event.

In the case of $^{22}$Na, we use coincidence summing to distinguish between events originating from $β^+$ and EC branches: The addition of one or both annihilation quanta to the response from the 1275 keV de-excitation photon makes that the upper part of the spectrum $(>1275\,keV)$ represents $β^+$ decay only. Lower sections of the spectrum, on the other hand, are fed by both branches: the full energy peak of the de-excitation photon and its Compton tail following electron capture, plus partial absorption of the de-excitation and annihilation quanta following $β^+$ decay.

### E. Data acquisition and analysis

The chain of data acquisition and analysis is as follows:
1. A pulse-height spectrum is built during a 2-hour livetime period. Its completion automatically triggers the start of the next.
2. After a long period of unattended operation, spectrum files are retrieved from the data acquisition laptop via local access.
3. The spectra are calibrated to a 0–3 MeV scale, see Sec. II-F.
4. RoIs are set to achieve maximum power of distinction between different modes of decay if they exist, and their countrates are calculated. For the $^{22}$Na source measurements we define the RoIs TOT, MED, and HI.
5. Background countrates in these RoIs are subtracted. The manner in which background is assessed is discussed in Sec. III.
6. The resulting net countrates are corrected for natural decay of the used source since a fixed date. Although not strictly required, this facilitates graphing changes in countrates over long data taking periods, e.g. for the purpose of outlier rejection.
7. The resulting decay-corrected countrates from each 2-hour spectrum are combined to daily averages. Although not strictly required, this facilitates graphing as it averages out Poisson counting scatter to a significant degree.
8. These daily averages are used to fit a function that describes all known phenomena. It not only contains a parameter for the relative in- or decrease in countrates when the reactor changes status but also incorporates parametrizations for a number of instrumental effects associated with the detector setup. These artefacts are discussed in Sec. IV.
9. Using the standard methodology of least-squares fitting, the parameter uncertainties and the goodness of fit are determined by using Poisson variances as weighting factors and by testing $χ^2$ on the basis of its right-tail probability P (p-value). For $P>0.05$, the fit is accepted. For $P≤0.05$, either the fit model assumptions are rejected or, if no obvious cause of the poor fit can be found, the internal variances of the deduced parameters are multiplied by the reduced $χ^2$ to provide a best estimate of their statistical (random) uncertainty.

### F. Energy calibration

During acquisition, spectra are pulse-height stabilized by online fine-adjustment of the PMT gain, using a preselected dominant peak in the spectrum. (This is also known as 'gain locking'.) As discussed in Sec. II-C, further offline analysis is needed to convert pulse-height histograms into energy spectra with suitable long-term peak position and RoI stability. A lack of such stability can affect stability in countrate measurement over long periods ($\square\,1\,week$) of data taking.

A two-stage approach is followed. First, a number of peaks known to be prominent in the raw spectra are localized. Second, a map between channel locations and true energies is calculated and the pulse-height distributions are converted to calibrated energy spectra. Our pulse-height spectra resemble a number of peaks residing on an exponentially declining continuum. The γ-ray lines used to stabilize and calibrate the spectra are at 511, 1022, 1275, 1786, and 2297 keV for spectra taken with the $^{22}$Na source, at 811, 1173, 1333, and 2506 keV for $^{60}$Co spectra, and at 511, 1461, and 2614 keV for background spectra acquired without any of the sources. The following peak stabilization procedure was applied:

1. Raw spectra were smoothed.
2. In a wide energy range around the peaks, parts of the spectra were plotted on a semi-logarithmic scale and the linear continuum (exponential in the original spectrum) was subtracted to make the peaks stand out more prominently over the continuum.
3. With a search algorithm the potential position of the peak is determined at the nearest channel number.
4. The peak location is refined by least-squares fitting a second-order polynomial (corresponding to a Gaussian in the original spectrum) to the top of the peak. This implies a fitting range not wider than the peak FWHM. This FWHM is estimated from the following assumed energy-dependency and pivot point (7% at 662 keV):

$$\frac{FWHM}{E} = 0.07 \sqrt{\frac{E}{662\,keV}}\,. \qquad (1)$$

5. The peak locations in fractional channel number were used to construct a linear energy scale with a range of 0 to 3 MeV. This step involves the redistribution of the counts according to the method described in [22,23].

III. BACKGROUND ASSESSMENT

A. General considerations

Research reactors are constructed to facilitate experiments at close proximity to the reactor core. This implies that there are penetrations in shielding walls and caves with neutron beams. Wall thicknesses, often $< 2\,m$, are mainly chosen to satisfy radiation safety regulations and without the objective of low background in mind. Recently, an investigation of the background at three USA research reactors was published [20]: Oak Ridge National Laboratory (ORNL), National Institute of Standard and Technologies (NIST), and Idaho National Laboratory (INL). In preparation for short-range antineutrino oscillation experiment PROSPECT [24], the authors thoroughly mapped background radiation at possible locations for their detector near these reactors. They investigated the presence of slow and fast neutrons, including cosmogenic neutrons, γ-rays, and muons. The results differ considerably between the three facilities and among the locations within each facility. The main cause for these differences is traced back to openings and feedthroughs in the shielding walls, for pipes, ducts, and beamlines. Higher site elevation, associated with less atmospheric attenuation of cosmic ray secondary particles, is found to be a relatively small factor amongst the background radiation sources at these three sites.

Maximizing measurement sensitivity for any effect of antineutrinos on the decay rate of β-unstable nuclei requires many factors to be optimized or chosen correctly. The choice to position the detector setup in the seismic vault under a ca. 8 m thick solid slab of concrete ensures that background is only a minor fraction of the detector's total response to a sufficiently strong source nearby or inside. This reduces the requirement to identify and monitor individual sources of background, for example with a high energy-resolution detector. This situation is in stark contrast with reactor antineutrino detector systems based on the IBD detection principle where background countrate is a significant if not dominant fraction of the total event rate. Cosmogenic contributions are not only reduced by the reactor and the reactor building at Koeberg but also by the fact that the detector is at 3 m below sea level.

B. Background measurements

We do not employ a duplicate detector setup to simultaneously measure the decay of the beta source and background radiation. Instead, background is measured with the same NaI well detector and shielding setup as used for the decay measurement but at different times. Keeping in mind the objective and nature of our experiment, see above, the relative change in background during ON and OFF reactor states is more important to quantify and to account for than determining its absolute value with the highest possible accuracy.

Background during reactor ON and OFF states was assessed in 2015 with measurement periods of 334 and 634 hours, respectively. The energy spectrum during the OFF period is shown in Fig. 2, scaled down by a factor 20 to highlight (dis)similarities with the unscaled ON–OFF difference spectrum. It is expected that the background in the seismic vault is predominantly from γ-rays emitted in the decay of $^{40}$K and nuclei in the decay series of $^{235}$U, $^{238}$U and $^{232}$Th present in the concrete. We identify the 511 peak from positron annihilation, the 2614 keV peak from the $^{232}$Th series and the 1460 keV peak from $^{40}$K. The 7.5 cm thick Pb shielding around the NaI detector absorbs low-energy γ-rays more strongly than the higher energy ones. A small complication is that the energy resolutions of the average ON and OFF spectra differ (because of difficulties with the online gain stabilization feature), adding some artificial structure to the difference spectrum.

Still, it is clear that the 2614 keV peak is enhanced by a change from reactor OFF to ON and more so than the relative enhancement of the 511 keV peak. This observation cannot be explained by incomplete subtraction of the background during reactor OFF status from the background during ON. According to [20], fast neutrons may be responsible for the main difference between the background spectra during the ON and OFF status. Fast neutrons can excite $^{208}$Pb, present in the lead shielding, to its first excited state at 2.614 MeV. If this takes place in our setup, then it is promptly followed by emission of the 2614 keV γ-ray for de-excitation back to the groundstate. Note that this mechanism cannot produce additional γ-rays from other nuclei such as the 911 keV γ-ray ($^{228}$Ac of $^{232}$Th decay series). The long absorption path length of fast neutrons in Pb will cause these de-excitation γ-rays to be produced (near-)uniformly over the volume of the shielding, reducing the effective thickness of the lead shielding for absorbing these particular γ-rays. Our observation of background is consistent with decay of actinides and daughter products in the concrete of the seismic vault and with the presence of fast neutrons in the seismic vault when the reactor is ON.

Figure 3 shows a typical pattern of variability in the background countrate. The daily-averaged data shown are the result of integration over a substantial part of the energy spectrum and span a 26-day counting period in 2015 when the reactor was OFF. The scatter around the average value ($\chi^2_{red} = 3.3$) is more than expected from the Poisson counting process ($P = 0.05$ corresponding to $\chi^2_{red} = 1.5$). Since there is no obvious cause for the larger scatter the external uncertainty in the average value is used in the propagation of uncertainties.

After a careful matching of the energy calibrations of the ON and OFF background spectra the ON/OFF ratio values for the RoIs TOT, MED, and HI were obtained as 1.0265 ± 0.0006, 1.0275 ± 0.0012, and 1.0342 ± 0.0008, respectively. These factors were used to deduce the background during the OFF periods from the measured ON values in the periods Dec-2012–May-2013 and Oct-2013–Feb-2014.

## IV. $^{22}$Na MEASUREMENTS

### A. General considerations

A typical spectrum for $^{22}$Na inside our well counter is shown in Fig. 4. In addition to two expected peaks from the single-photon lines at 511 and 1275 keV, coincidence summing peaks appear at $(2 \times 511 = 1022)\,keV$, at $(1275 + 511 = 1786)\,keV$, and at $(1275 + 2 \times 511 = 2297)\,keV$. To understand the main features, we compare in the figure the shape of the measured spectrum with one constructed from simulating the fate of the $^{22}$Na decay characteristic photons in our detector setup, using the simulation code MCNPX. The details of these simulations are found in the Appendix. The structures in the measured continuum are reproduced reasonably well. The absolute magnitude, especially at lower energies, depends strongly on the details of the source geometry. Relative to $^{22}$Na spectra recorded with common NaI crystals, the continua under and between the peaks appear reduced. Clearly, the well geometry makes it more probable that energy depositions by Compton scattering, normally responsible for continua on the low-energy side of peaks, sum towards higher energy values. This implies that all $^{22}$Na events in the spectrum with an energy higher than the 1275 keV line are only due to summed $β^+$ signals: detection of one or both annihilation quanta in true coincidence with the 1275 keV deexcitation photon.

### B. Specific choices for $^{22}$Na

Three RoIs are defined. Their labels and ranges are: TOT 170–2452 keV, MED 1151–1351 keV, and HI 1353–2452 keV. HI exclusively contains pulses from $\beta^+$ decay, whereas MED captures both EC and $\beta^+$ decay. While TOT has maximum overlap with MED and HI, it also covers the two annihilation peaks and their continua. Considering also the branching ratios of $^{22}$Na decay, TOT is thus dominated by $\beta^+$ decay. The rationale for coincidence summing is now clear: Without it, distinguishing between the two decay modes with a γ-ray detector is more difficult as it requires accurate knowledge of the 511 and 1275 keV peak shape and efficiencies.

After subtracting background (BG) from the gross $^{22}$Na (Na+BG) countrates, for each of the RoI separately, the net countrates were corrected for natural decay with the literature decay constant $\lambda = 3.039 \times 10^{-5} \, h^{-1}$ [25]. The activity of the $^{22}$Na source was 2 kBq in November 2012.

The first measurement started on 12-Dec-2012 and the last one ended on 2-Dec-2015. From these, we selected two continuous measurement series, each comprising one ON to OFF plus one OFF to ON reactor status change. The calendar periods are given in Table II. While detector and electronics were left untouched during each series, the $^{22}$Na source was removed and put back once during the first series for the purpose of a BG measurement.

### C. Unexpected artifact in acquisition livetime

After averaging the 2-hour long spectra to 1-day spectra, an unexpected structure with a 1-month periodicity became visible. In Fig. 5 we show this daily countrate for TOT during the first and second measurement series. The moments at which countrates instantaneously dip by ca. 0.05% unequivocally point to one root cause. At noon on the first day of each month during the first measurement period, a rescue and recovery utility (process LaunchRnR) on the data acquisition computer had been activated inadvertently. Our most plausible explanation is that the associated increase in CPU load interfered with the livetime bookkeeping by the Venus pulse-height digitization electronics and/or the Interwinner acquisition software. Further evidence for the identity of this effect comes from the observation that its occurrence shifts in time by an amount that is exactly equal to the shift in the date-time set on the laptop by the user. Moreover, the dip structure does not reappear if the rescue and recovery utility is turned off.

The second measurement series was aimed at testing repeatability of the first. To not change anything in the setup, we chose to leave the rescue and recovery utility on. In the right panel of Fig. 5 the dip structures show up consistently. In comparing the two panels one notices that the dip pattern and the ON and OFF periods do not coincide. Moreover, one notices a trend in which the counts during ON are below the solid line and during OFF above that line indicating a reactor-status dependence of the decay corrected countrates.

### D. Description of $^{22}$Na countrate data

Obviously, the presence of these artificial dips can severely hamper any reliable identification and quantification of a possible link between $^{22}$Na decay rate and reactor status. However, it appears that the time dependence and the magnitude of these dips are repeatable and describable by exponential decay, see Fig. 5. This possibility led us to attempt to search for a parametric description of each countrates series (as stated above: natural-decay corrected, 1-day integrated, and scaled to near unity), with parameter values to be found through least-squares regression.

We describe the two countrate time series by assuming the presence of three effects:
1. the dip structure,
2. a reactor-status dependence, and
3. a residual trend with time.

The first effect is described by rapid exponential decay (half-life $\approx 15 \, days$) of a small ($\leq 0.1\%$) excursion at known times. Knowing that the dips are caused by an instrumental effect on the digital side of the signal chain, we can safely assume that its presence is uniform across the energy spectrum. Hence, the magnitude of the dip structure should be the same for any RoI. This simplifies the task of disentangling the various effects. The third effect is characterized by a small and linear ($< 3 \times 10^{-5} \, day^{-1}$) decrease in countrate across each measurement period. We attribute this effect to the possibility that pile-up and baseline shift are compensated for only to first order in the conversion of pulse-height to energy spectra (see Sec. II-F). If correct, the trend rate will likely vary amongst the RoIs. Another possible explanation for this global trend is having chosen an incorrect value of the $^{22}$Na decay constant in the decay correction step. Effect #2 is described as a fractional change in countrate when the reactor is ON relative to when it is OFF. Allowing for the possibility that the two $^{22}$Na decay modes are influenced differently by reactor-status changes, we reserve a fractional change parameter for each RoI separately.

Each RoI countrate series receives its own description of all three time-dependent effects mentioned above. Effect #1 is determined from the TOT series only because the effect is pulse-height independent and because that RoI has the highest Poisson precision. Introducing the independent time-variable $d$ for the number of days since 31-Dec-2012, we factorize the three effects in the regression fit function as follows:

$$CountRate(d, RoI) = dip(d)\, react(d, RoI)\, trend(d, RoI)\, aver(RoI) \tag{2a}$$

where the first three terms on the RHS are defined by:

$$dip(d) = 1 + dip_{ampl} \exp\{-dip_{decay}[d - onset(d)]\}, \tag{2b}$$

$$react(d, RoI) = 1 + frac(RoI)\, status(d), \tag{2c}$$

$$trend(d, RoI) = 1 + trend_{rate}(RoI)[d - pivot]. \tag{2d}$$

The function $onset(d)$ is the value $d$ of when the most recent dip was triggered. The function $status(d)$ is binary: $status(d) = 1$ at times $d$ when reactor status is ON and $status(d) = 0$ when it is OFF. Nuisance parameter $aver(RoI)$, the fourth term in Eq. (2a), is needed because observable $CountRate(d, RoI)$ was scaled to a value $\approx 1$ prior to the regression. Parameter $pivot$ is fixed before the regression. By making it equal to the time-midpoint of the data series, unnecessary correlation between $aver(RoI)$ and $trend_{rate}(RoI)$ is avoided. Equation (2) is our parametrization to isolate the reactor-status dependence from all other systematic effects that we have been able to identify.

### E. $^{22}$Na measurement results

We present in Table III the regression results for the two measurement series with the $^{22}$Na source. Each series was fitted twice: Once to test the full parametrization as given in Eq. (2) and once with the reactor status ignored, i.e. by modifying Eq. (2c) to

$$react(d, RoI) = 1. \tag{2e}$$

As is obvious from the $\chi^2$ test statistic and its associated p-value, the inclusion of the reactor-status step function very significantly improves the description of the measured countrate data in both series. The second observation is that the two series replicate each other with regards to all three reactor effect fractions $frac(RoI)$. Furthermore, fraction $frac(MED)$ is slightly positive, indicating an increase in countrate in the interval 1151–1351 keV when the reactor is ON, relative to when it is OFF. Fractions $frac(HI)$ and $frac(TOT)$, on the other hand, are negative and by similar amounts. By rearranging Eq. (2a), the reactor effect for the three RoIs in the two measurement series can be shown graphically, see Fig. 6:

$$react(d, RoI) = \frac{CountRate(d, RoI)}{dip(d)\, trend(d, RoI)\, aver(RoI)}. \tag{3}$$

Countrate data obtained during the transition periods T, in which the antineutrino flux is changing due to the cool-off of the fuel or the ramp-up of the reactor, have not been used in the regression fits.

In Table IV we summarize the fractional countrate change values due to the reactor effect, found through regression of the two measurement series with the $^{22}$Na source, for each RoI separately. The quoted statistical errors are external uncertainties, thus Poisson uncertainties multiplied by $\left(\chi^2/dof\right)^{1/2}$ if warranted by $P < 0.05$.

Table IV also provides averages of those two measurement series, including their statistical and systematical uncertainty, for each RoI. The limited precision and accuracy in the ratio of background countrate during reactor ON relative to that during reactor OFF are assumed to be the only source of systematical uncertainty in the quoted fractional countrate changes. Because the continuum in the background has a roughly exponential energy dependence (see Fig. 2) while in the $^{22}$Na spectrum the continuum is almost linear (see Fig. 4), the importance of this systematic uncertainty decreases with increasing energy. It is clear that our methods to suppress, determine, and correct for background radiation yield $\Delta A/A$ values with uncertainty dominated by statistical error.

As explained, RoI HI has no sensitivity for EC, and thus countrate changes in that RoI exclusively correspond to $\beta^+$ decay rate changes. The reactor-status effect on its $\Delta A/A$ value in each series measurement differs from zero by $> 5\sigma$ and by $\approx 10\sigma$ in the weighted average. From these observations we draw the tentative conclusion that $\beta^+$ decay of $^{22}$Na is suppressed when the reactor is ON.

We note that the antineutrino flux decreases during the burn-up cycle ($\approx 545\,days$) by about 2%, see Table I. Measurement series 1 and 2 are ca. 240 days apart (see Table II), a period during which no refueling has taken place. If the reactor-status effect is proportional to the antineutrino flux, then the decrease in that effect during that period would be about 0.7%. Our analysis shows that the fractional countrate change values of RoI TOT and HI decrease from series 1 to 2 (see Table IV), but without statistical significance: In relative terms, $frac_{TOT}$ reduces by $[4\pm9\text{(stat)}]\%$ and $frac_{HI}$ by $[16\pm14\text{(stat)}]\%$. Hence, it is obvious that our present sensitivity is insufficient to assess fuel burn-up at the 1% level.

## V. $^{60}$Co MEASUREMENTS

### A. Motivation

The $^{60}$Co groundstate ($J^\pi = 5^+$) has only one mode of decay: $\beta^-$ decay to predominantly the $E_x = 2506\,keV$ state of $^{60}$Ni ($J^\pi = 4^+$). Electron capture and $\beta^+$ emission to the groundstate of $^{60}$Fe $J^\pi = 0^+$ have negative Q-values and are thus six-times forbidden. Following $\beta^-$ decay of $^{60}$Co, the dominant de-excitation mode to the $^{60}$Ni groundstate is cascaded emission of two γ-rays ($E_\gamma = 1173, 1333\,keV$). Their energy is comparable to the $E_\gamma = 1275\,keV$ γ-ray in the $^{22}$Na decay. They will also lead to coincidence summing with our source-detector geometry. Hence, purely in terms of energy and coincidence timing of emitted photons, the $^{22}$Na and $^{60}$Co sources will lead to similar responses in our detector setup. If the reactor-status effect observed with the $^{22}$Na source is not seen with a $^{60}$Co source, then the occurrence of the effect is associated with $\beta^+$ decay. If, on the other hand, the reactor-status effect shows up in $^{60}$Co as well, then its explanation lies most probably in the design, execution, and/or analysis of the experiment.

### B. Measurement

The $^{60}$Co measurements were conducted with a similar source strength as the $^{22}$Na source. Apart from the absence of the MCA livetime dips and the use of an analog MCA, the measurement conditions with these two sources were very similar. In Fig. 7 we show the decay-corrected countrate in the range $E_\gamma = 2376–2719\,keV$ during the 23-Dec-2014 to 5-May-2015 period, including a reactor outage. The reactor ON–OFF effect for these $^{60}$Co data is $\Delta A/A = [-2\pm6\text{(stat)}]\times10^{-5}$ ($\chi^2_{red} = 0.93$). This value is an order of magnitude lower than the values obtained for RoI HI in the $^{22}$Na case and shows that if the reactor-status effect for $^{60}$Co is non-zero then it is below our detection limit. We conclude from the absence of a measurable reactor-status effect with $^{60}$Co that there is no simple instrumental or procedural explanation for the highly significant reactor-status effect observed with $^{22}$Na (see Sec. IV-E and Table IV).

## VI. $^{50}$V MEASUREMENTS

### A. Motivation

The $^{50}$V groundstate ($J^\pi = 6^+$) decays with a half-life of 1.4 x $10^{17}$ y by electron capture (83%) to $^{50}$Ti(1) at $E_x = 1554\,keV$ ($J^\pi = 2^+$) and by $\beta^-$ decay (17%) to $^{50}$Cr(1) at $E_x = 783\,keV$ ($J^\pi = 2^+$) [25]. The decay to $^{50}$Ti(1) is purely EC since $Q(\beta^+) = -368\,keV$, see the decay scheme in Fig. 8. Capture of sufficiently energetic antineutrinos to overcome this threshold, i.e.

$$^{50}_{23}V + \bar{\nu}_e \rightarrow {^{50}_{22}Ti(1)} + e^+ \tag{4}$$

can be tested with a $^{50}$V source in our detector setup at the Koeberg reactor since the bulk of the reactor neutrino spectrum lies above $E_{\bar{\nu}} = 368\,keV$. However, in contrast to the well-known IBD reaction on the free proton, the above capture reaction is strongly hindered by the required change in nuclear spin $J$.

## B. Measurement

In an attempt to compensate for the expected low cross section of this capture process, we placed in our well counter a 27.6 g rod of natural vanadium (0.25% $^{50}$V), corresponding to 8.3 x 10$^{20}$ $^{50}$V nuclei and 0.13 mBq of natural activity. In the period 21-Aug–17-Oct 2015 we acquired spectra with this vanadium rod while the reactor was ON and from 18-Nov to 2-Dec 2015 without the V rod, to provide the reactor ON background. To optimize sensitivity in detecting γ-rays related to antineutrino capture on $^{50}$V three RoIs were set. The average countrate for RoI 320–3000 keV with the V rod was 40539 ± 6 h$^{-1}$ and without it 40601 ± 11 h$^{-1}$. Similar countrate reductions with the V rod in place were observed for RoIs 794–1614 and 1617–3000 keV and are likely due to absorption of background γ-rays by the V rod. Hence, we must conclude antineutrino capture in $^{50}$V has not been observed.

## VII. DISCUSSION

In this section, we discuss the robustness of our measurement data, attempt to express in them in terms of experimental cross sections, and offer pathways for interpretation.

### A. Consistency check

We measured a non-zero reactor-status effect for $^{22}$Na in the countrate of three γ-ray energy intervals (see Sec. IV-E and Table IV). If its cause is physical then it is obvious to try to express its magnitude in terms of changes in decay constants. Moreover, since we have derived three parameters while $^{22}$Na has only two dominant decay modes, we can perform a check of the simple model that our three observed countrate changes can be described by two decay constant changes. A lack of consistency can mean two things: Either our measurement data are plagued by systematic effects that we have not been able to identify and quantify, or the model that the reactor-status effect for $^{22}$Na can be expressed in terms of only two decay constant changes is too simple.

From the basic decay law

$$A = \lambda\,N, \tag{5}$$

where $A$ is the activity, $\lambda$ the decay constant, and $N$ the number of nuclei, it follows that the fractional countrate change ratio must equate the fractional change in the decay constant associated with the registered events:

$$R \equiv \frac{\Delta A}{A} = \frac{\Delta \lambda}{\lambda}. \tag{6}$$

When specified for particular decay mode $i$, changes in $\lambda_i$ may alter its branching ratio

$$BR(i) = \frac{\lambda_i}{\lambda}. \tag{7}$$

Since the $\beta^+(0)$ transition to the $^{22}$Ne ground-state is very weak ($BR < 0.1\%$), we ignore it here and limit the check between EC and $\beta^+$ decay to the $^{22}$Ne(1) state at $E_x = 1.275$ MeV. Thus, with abbreviated notation for the $\beta^+(1)$ decay constant, we get:

$$\lambda = \lambda_{EC} + \lambda_{\beta^+(0)} + \lambda_{\beta^+(1)} \approx \lambda_{EC} + \lambda_{\beta^+(1)} = \lambda_{EC} + \lambda_\beta. \tag{8}$$

As explained in Sec. IV-B, RoI HI can only be populated by coincidence summing involving $\beta^+$ decay, thus

$$R_{HI} = \frac{\Delta\lambda_\beta}{\lambda_\beta}. \tag{9}$$

Under the assumption that the reactor status has no effect on EC, i.e. $\Delta\lambda_{EC} = 0$, we can express the ratio between the relative countrate changes in RoIs TOT and HI as an algebraic term involving two ratios:

$$\frac{R_{TOT}}{R_{HI}} = \left[1 + \left(\frac{\eta_{TOT,EC}}{\eta_{TOT,\beta}}\right)\left(\frac{\lambda_{EC}}{\lambda_\beta}\right)\right]^{-1}. \tag{10}$$

The first ratio has in the numerator the detection efficiency of electron capture events being registered in RoI TOT and in the denominator the detection efficiency that $\beta^+$ decay events are registered in that same RoI. The second ratio is that of the unmodified decay constants. The two detection efficiency values have been determined from the Monte Carlo simulation of the source-detector setup (see the Appendix and Fig. 9): $\eta_{TOT,EC} = 0.557$ and $\eta_{TOT,\beta} = 0.954$. The RHS of Eq. 10 then becomes $0.941$. Using the experimentally determined values in Table IV, the value for the LHS is $1.12 \pm 0.12(stat)$. The two values are is not in gross contradiction with each other.

More generally, the relative countrate changes in RoIs TOT and HI can, in combination with the known branching ratios, be used to estimate the fractional change in the EC decay constant:

$$\frac{\Delta\lambda_{EC}}{\lambda_{EC}} = R_{TOT} + \left(R_{TOT} - R_{HI}\right)\left(\frac{\eta_{TOT,\beta}}{\eta_{TOT,EC}}\right)\left(\frac{\lambda_\beta}{\lambda_{EC}}\right) = \left[-8 \pm 5(stat)\right]\times 10^{-4}. \tag{11}$$

This value appears plausible but leads to an inconsistency. From the MC simulation it is estimated that RoI MED is for 26% fed by EC and for 74% by $\beta^+$ emission. It is now obvious that, regardless of the accuracy of these percentages, it is impossible to combine the negative $R_{TOT,HI}$ and $\Delta\lambda$ values, as per Eqs. 9 and 11, to the positive value for $R_{MED}$. We have no explanation for this particular lack of consistency between the reactor OFF and ON spectra for $^{22}$Na. Possibly, the appearance of the reactor-status effect in the γ-ray energy spectrum of this nuclide is more complex than just modification of the two largest decay constants. In Sec. VII-D we discuss thermal neutron absorption on $^{22}$Na and subsequent proton decay to $^{22}$Ne(1) as a possible explanation.

### B. Sources of systematic uncertainty

The main goal of this investigation is to assess if a reactor-status effect is observable in γ-ray countrates associated with particular modes of β-decay. The effect that we have observed for $\beta^+$ decay of $^{22}$Na is small and thus must be tested for robustness against random and systematic errors. In Sec. VII-C we will attempt to express our countrate results as antineutrino interaction cross sections. Obviously, that calculation requires knowledge of the flux of incoming reactor neutrinos. As discussed in Sec. II-B, our flux calculation has a 30% systematic uncertainty. Note however that this large error has no impact on the size of the reactor-status effect.

Quantifying relative countrate changes at the $\leq 10^{-4}$ level not only requires a high number of counts but also long-term stability in detection efficiency and in time measurement, accurate energy calibration, and accurate background countrate assessment. The $^{22}$Na and $^{60}$Co measurements were processed in the steps discussed in Sec. II-E. After energy calibration and background subtraction, the 2-hourly countrates were corrected for natural decay, then averaged to daily countrates, and finally parameterized in terms of instrumental and reactor-status effects.

Good calibration of pulse-height spectra is necessary to achieve the required stability in countrates. It is, *a priori*, not inconceivable that the reactor status somehow influences the photomultiplier gain and/or pedestal. Although these effects have been mitigated by the use of a DSP-based MCA and online gain stabilization, we have seen that offline calibration was needed for each spectrum separately, to reduce scatter in the location of full-energy peak positions and RoI boundaries. No link was seen between energy calibration constants and reactor status.

Only for perfect shielding one may expect that no reactor-status changes are present in the background radiation levels. While placement of our setup in the seismic vault gives a very large suppression of radiation from the reactor core and its immediate surroundings, it is clear that even the slightest change in background radiation, if not accounted for, can mimic the effect we want to quantify. The large difference in spectral shape between background and source measurements make this particularly true for RoIs covering the low-energy part of the spectrum. As presented in Sec. III, we assessed background with the same NaI crystal and Pb castle as for the $^{22}$Na, $^{60}$Co, and $^{50}$V measurements. The inherent disadvantage of that choice is that background cannot be assessed simultaneous with source measurements. We have quantified the difference in background between reactor ON and OFF and explain it as inelastic scattering of fast neutrons, present during reactor ON only, off $^{208}$Pb nuclei in the Pb castle. Another detail that deserves further attention is whether changes in the effective attenuation of fast cosmic-ray neutrons with (un)loading of the reactor have any effect on the background spectrum. For each RoI separately, the background-countrate precision has been propagated as a source of systematic uncertainty in the fractional countrate ratio changes defined by Eq. (6). The results obtained with the $^{22}$Na source (Table IV) show that for all three RoIs the systematic uncertainty is smaller than the statistical uncertainty.

The occurrence of countrate dips due to poor online livetime determination during acquisition of the $^{22}$Na data was discussed in Sec. IV-C. This problem was surmounted by identifying the root cause, quantifying its effect, and by incorporating it in a least-squares regression of all effects simultaneously. From Figs. 5 and 6 (see Secs. IV-D,E) it is clear that not only have the livetime dips been effectively accounted for but also that, if they had not been, the correlation between $^{22}$Na countrate and reactor-status effect would still have been visible. Hence, the livetime dips do not add to the total systematic uncertainty in the relative $^{22}$Na countrate changes when the reactor changes between the ON and OFF states.

Another aspect of livetime determination is frequency stability of the clock-oscillator used by the FPGA circuit for DSP functions in the MCA. Because we have not been able to identify the model and make of the clock-oscillator, we do not know its specifications such as temperature dependence. However, if we assume a worst-case scenario of 100 ppm variability in the frequency, which for modern oscillators would require large swings in ambient temperature, then it can explain our observed reactor-status effect only if those swings coincide with reactor-status changes. Since the air temperature at the location of the setup remains constant within 2°C (see Sec. II-A), this explanation seems unlikely. It is conceivable that other factors such as fluctuations in mains voltage and ambient humidity (neither of those were monitored) have affected the clock-oscillator's performance. In hindsight the MCA's livetime determination should have been tested before the experiment with regards to long-term repeatability and more effort should have been put into monitoring of environmental parameters.

Errors in the chosen natural half-lives for $^{22}$Na and $^{60}$Co and in the corrections for pulse pile-up and baseline shift (see Sec. II-C) are noticeable only over significant fractions of a half-life. They cannot influence the quantification of relative countrate changes on a much shorter timescale. Moreover, we have effectively captured those errors as a slow linear trend in the countrate parameterizations.

We have failed to find a mechanism by which artefacts or shortcomings of instrumental or data analysis nature can explain the non-zero reactor-status effect observed with the $^{22}$Na source. Most compelling is that we see no such effect with $^{60}$Co in the same setup.

### C. Derivation of cross sections

Here, we make explicit use of the assumption that the observed relative countrate differences ( $\Delta A/A$ ) between reactor ON and OFF are due to a flux $\phi_{\bar{\nu}}$ of antineutrinos. In the definition of total interaction cross section $\sigma$, we equate the rate of target interactions $I$ with the change in source activity $\Delta A$:

$$\sigma \equiv \frac{I}{\phi N} = \frac{|\Delta A|}{\phi_{\bar{\nu}} N} = \frac{\lambda}{\phi_{\bar{\nu}}} \frac{|\Delta A|}{A}, \qquad (12)$$

where $N$, the number of target atoms, is set to the number of source nuclei. Taking the absolute of the activity change as the number of target interactions is justified because the neutrino–nucleus interaction may lead to faster or slower decay in an existing decay mode.

The decay constants for $^{22}$Na and $^{60}$Co are 8.442 x 10$^{-9}$ s$^{-1}$ and 4.167 x 10$^{-9}$ s$^{-1}$, respectively, based on the half-lives listed in [25]. With fractional countrate changes $(\Delta A/A)_{TOT} = (-3.02 \pm 0.14) \times 10^{-4}$ for $^{22}$Na (see Table IV) and $\Delta A/A = (-2 \pm 6) \times 10^{-5}$ for $^{60}$Co (see Sec. V-B) and an antineutrino flux of $\phi_{\bar{\nu}} = 1.65 \times 10^{13} cm^{-2} s^{-1}$ (see Sec. II-B), the corresponding cross sections with their statistical uncertainty become (1.55 ± 0.07) x 10$^{-25}$ cm$^2$ for EC+$\beta^+$ decay in $^{22}$Na and (0.5 ± 1.5) x 10$^{-26}$ cm$^2$ for $\beta^-$ decay of $^{60}$Co. As mentioned in Sec. V-B, the value for $^{60}$Co reflects a detection limit. For $^{22}$Na the value is well above the detection limit and exceeds the well-known value of 10$^{-43}$ cm$^2$ for low-energy antineutrino capture by a free proton by 18 orders of magnitude.

### D. Neutron capture by $^{22}$Na

The nuclide $^{22}$Na has a very large cross section for (thermal) neutron capture: 2.83 x 10$^{-20}$ cm$^2$ [26]. Since the dominant decay mode of the $^{23}$Na capture state is proton decay to $^{22}$Ne(1) [27], it is followed by the same 1275 keV γ-ray around which we have defined RoI MED. So, this process can lead to a countrate increase in RoIs MED and TOT when the reactor changes its state from OFF to ON, but not in RoI HI. There are two reasons why our observations cannot be explained by thermal neutron capture. First, we observe a negative change in countrate in RoI TOT in during transitions from OFF to ON. The second argument follows from an estimation of the neutron flux required to cause the magnitude of the observed effect. Scaling down the reactor neutrino flux by the ratio between the abovementioned neutron-capture cross section and the cross section we derived for influencing the total decay rate of $^{22}$Na (see Sec. VII-C), we arrive at a (thermal) neutron flux equal to 10$^8$ cm$^{-2}$ s$^{-1}$, which corresponds to a dose rate of 140 Sv h$^{-1}$. However, the seismic vault area in which we have placed our setup is outside the radiation zone (see Sec. II-A), meaning that the dose rate should be less than 1 mSv y$^{-1}$ [28].

### E. New physics?

Low-energy electron (anti)Neutrino Capture on Beta decaying nuclei (NCB) was studied by Cocco et al. [14] within the framework of nuclear structure and Fermi theory of β-decay. With future application to measuring cosmological relic neutrinos in mind, for which low backgrounds are an absolute necessity, they searched the entire ENDSF database for parent-daughter transitions yielding branching ratios > 80% and the highest product of cross section, neutrino velocity, and half-life. Amongst the $\beta^+$ decaying ones found is $^{22}$Na, with $\sigma_{NCB}(v_\nu/c) = 3.04 \times 10^{-48} cm^2$. Other $\beta^+$ unstable nuclides in that group ($^{18}$F, $^{45}$Ti) have cross sections that are 3–4 orders of magnitude higher, but their half-lives are shorter by the same factor, thereby not improving the discrimination of NCB events against the process without capture. Amongst the super-allowed $0^+ \rightarrow 0^+$ transitions, the highest cross section found was for $^{54}$Co, with $\sigma_{NCB}(v_\nu/c) = 1.2 \times 10^{-42} cm^2$ and partial half-life 0.2 s.

We derived a cross section of $\sigma_{exp} = 1.6 \times 10^{-25} cm^2$ from the reduction of the γ-countrate uniquely identifying $\beta^+$ decay of $^{22}$Na when the Koeberg reactor changes its status from OFF to ON. The 17 orders of magnitude gap between prediction and measurement suggests that the measurement response in our experiment is due to a process other than neutrino capture and/or that the study by Cocco et al. [14] has incorrectly identified the leading mechanism(s) at low neutrino energies.

Possibly, the sign of the observed reactor effect on the $^{22}$Na $\beta^+$ decay rate yields information about the dominant neutrino interaction mechanism. Fermi's Golden Rule expresses the transition rate between initial and final eigenstates as the quantum-mechanical product of their respective wave function, the interaction (perturbation) Hamiltonian, and the density of final states. From that viewpoint, it is impossible to obtain a reduction of a $\beta^+$ decay constant by projectiles without some change in one or both wave functions. Following on from that, we speculate the existence of two mechanisms:
1. The initial state of $^{22}$Na is modified through an interaction e.g. inelastic scattering of the incoming electron-antineutrino on a proton bound in the nucleus. Inelastic scattering could change the spin orientation of a proton, e.g. from $l_j = d_{3/2}$ to $d_{5/2}$, and thereby slightly modifying the $^{22}$Na wave function, resulting in a slower decay rate. Such a process is highly unlikely for a free proton, being pure $s_{1/2}$.
2. The incoming electron-antineutrino couples to the emitted electron-neutrino in the final state, forming a quasi-bound state similar to that of an electron-positron pair in positronium. In our case, the quasi-bound state will be a neutral boson at an energy of less than 1.8 MeV, such that it cannot be formed with a free proton. On the one hand, according to the current knowledge of the weak interaction this would imply an additional factor $G_F^2$, with $G_F$ being the Fermi coupling constant $G_F/(\hbar c)^3 \approx 10^{-5} GeV^{-2}$, making the expected cross section almost negligible. On the other hand, recently a neutral boson has been put forward to explain observations of the electron-positron system created by internal pair conversion in the decay of $^8$Be(18.15 MeV) to the groundstate [29,30]. Because we see a reactor status effect for $^{22}$Na but not for $^{60}$Co this interpretation, if correct, may help to resolve the question whether neutrinos are Dirac or Majorana.

The above reasonings are qualitative only and are put forward to stimulate the search for explanations. They are not an attempt to explain the observed cross section for $^{22}$Na quantitatively.

## VIII. CONCLUSIONS

Jenkins et al. hypothesized that solar neutrinos influence the $\beta^-$ decay rates of $^{32}$Si and $^{226}$Ra, noticeable in the correlation between these decay rates and the annual variation of the Earth–Sun distance [15]. This triggered us to investigate whether a similar effect exists for $\beta^+$ decaying nuclei in the presence of reactor antineutrinos. In our first experiment, conducted at the pool research reactor at Delft, the Netherlands, no effect was seen for $^{22}$Na and the upper limit was an order of magnitude lower than the effect in $^{32}$Si at a comparable (anti)neutrino flux [16]. However, the measurements of that experiment were hampered by a high background mainly caused by decay of short-lived $^{41}$Ar produced by neutron activation of $^{40}$Ar in air. This led us to continue our investigation at a commercial nuclear power reactor with access to an area with a low and stable background. The seismic vault of the Koeberg nuclear power station fulfills these criteria with a three orders of magnitude higher flux difference between reactor ON and OFF than at Delft and a small difference in background between ON and OFF. The other major change was the deployment of a well-type NaI crystal rather than a cylindrical HPGe crystal, thereby not only increasing efficiency but also allowing discrimination between EC and $\beta^+$ decay modes through the use of coincidence summing. The recent work by Barnes et al. [19] sets an upper limit for an effect due to reactor status on the decay of $^{22}$Na: one or two parts in $10^4$. With a reactor neutrino flux of 3 x $10^{12}$ cm$^{-2}$ s$^{-1}$ at the HFIR (Oak Ridge, USA), their upper limit for the cross section is ca. 6 x $10^{-25}$ cm$^2$, which is roughly three times higher than the value we have determined.

In a previous version of this manuscript [17] no difference in background between ON and OFF was assumed, leading to a fractional (relative) change in the total $^{22}$Na γ-ray countrate between reactor ON and OFF equal to $[-0.51 \pm 0.11(stat)]$ x $10^{-4}$. Review and re-analysis of these same data focused on the following challenges:
- correcting higher order effects in online gain-stabilized energy spectra, through offline stabilization and calibration;
- problems with livetime measurement, brought under control by suitable parametrization and least-squares regression of countrate data within three RoIs; and
- subtracting different background countrates during reactor ON and OFF, and for the three RoIs separately.

The results of the present investigation can be summarized as follows:
1. For the widest RoI, capturing both EC and $\beta^+$ decay modes of $^{22}$Na, the fractional change equals $[-3.02 \pm 0.14(stat) \pm 0.07(syst)]$ x $10^{-4}$. Using an estimate of the antineutrino flux change between reactor ON and OFF, this result can be translated into a cross section equal to $[1.55 \pm 0.07(stat)]$ x $10^{-25}$ cm$^2$.
2. Combining the result from the widest RoI with that of the RoI that covers $\beta^+$ decay only suggests that the reactor-status effect for EC is similar to that for $\beta^+$ emission.
3. Taking also into account the observed fractional countrate change for the narrow RoI around the peak of the 1275 keV deexcitation γ-ray, leads to a numeric inconsistency that we cannot explain.
4. Replacement of the $^{22}$Na by a $^{60}$Co source results in a zero reactor-status effect.
5. While we have identified several shortcomings in the experiment, we have not been able to find an instrumental artifact that can explain the $^{22}$Na observations.

We tentatively interpret the $^{22}$Na results as a demonstration of a physics process of antineutrinos interacting with that particular nuclide or its constituents. If this process is antineutrino capture, then the study by Cocco et al. [14]

has incorrectly identified the leading mechanism(s) at low neutrino energies because their model predicts cross sections that are at least 17 orders of magnitude below what we have measured. We can also compare our findings for $^{22}$Na with the correlation between the Earth–Sun distance (R) and $\beta^-$ decay rates of $^{32}$Si and $^{226}$Ra as observed by Jenkins et al. [15]. Interpreting their 1/R$^2$ data as variations of the solar neutrino flux on Earth, their cross sections are four orders of magnitude higher than ours. In addition, their data show a relationship between flux and decay rate that is of opposite sign to ours. We therefore conclude that the hypothesis in [15] is either wrong or that the effect of solar neutrinos on $\beta^-$ decay of $^{32}$Si and $^{226}$Ra differs considerably from the effect of reactor antineutrinos on the $\beta^+$ decay of $^{22}$Na.

To summarize, there is a huge discrepancy between our observation of the reactor-status effect and current theoretical understanding of neutrino-nucleus interactions. Clearly our results with $^{22}$Na need to be replicated, with more emphasis on the influence from variations in background radiation, timing and energy stability, and environmental factors. Investigation with other β-decay nuclides will assist with theoretical model building and the design of future neutrino physics experiments. But we also hope that our findings will herald the development of compact neutrino detector systems for reactor physics research and strengthening of current non-proliferation measures.

## ACKNOWLEDGEMENTS

The authors gratefully acknowledge the Koeberg Nuclear Power Station for providing us access to the facility and Chr. Vermeulen at iThemba LABS for manufacturing the $^{22}$Na and $^{60}$Co sources. This work is financially supported by the Dutch Ministry of Foreign Affairs as part of the Member State Support Programme (MSSP) of the IAEA, by the Department of Physics and Astronomy at the University of the Western Cape, and from donations (money and in-kind) made to Stichting (Foundation) EARTH.

## APPENDIX

The simulated spectrum for $^{22}$Na inside our well-counter was shown in Fig. 4. It was created using MCNPX, version 2.5f [31]. Because this version does not handle coincidence summing between the $\beta^+$ emission to $^{22}$Ne(1) and the subsequent deexcitation within 3 ps [25], the following steps were taken:

1. Two independent, geometrically identical source-detector models are set up: one for $\beta^+$ emission by $^{22}$Na and one for emission of 1275 keV $^{22}$Ne deexcitation γ-rays. The emission rate for each of these particles is set to 1 Bq. The modelled source-crystal geometry is chosen equal to that of the experiment. The statistics and transport of scintillation light photons are not part of the model.
2. The models are run independently, each generating one histogram.
   - The positron annihilation probability spectrum contains the 511 and 1022 keV peaks and their Compton continuum. This spectrum is a convolution of the probability spectra $P^0_{e^+}$ and $P^1_{e^+}$ for the $\beta^+$ transition to the groundstate $^{22}$Ne(0) and to the first excited state $^{22}$Ne(1), respectively.
   - The γ-ray spectrum for the 1275 keV deexcitation of $^{22}$Ne(1) and its Compton continuum is denoted by $P_\gamma$.
3. The convolution takes the branching ratios $BR$ for the $\beta^+$ and EC channels into account. The final probability spectrum $P$ with histogram bin number $i$ is given by:

$$P(i) = BR_{\beta^+(1)} \left[ P^1_{e^+}(i)\left(1-\Sigma_\gamma\right) + P_\gamma(i)\left(1-\Sigma_{e^+}\right) + P_{CS}(i) \right] + BR_{\beta^+(0)}\, P^0_{e^+}(i) + BR_{EC}\, P_\gamma(i)$$

(A1)

$\Sigma_\gamma$ and $\Sigma_{e^+}$ are the summed probabilities for the detection of the 1275 keV γ-ray and the positron via annihilation radiation, respectively. $P_{CS}$ is the probability spectrum of true coincidence summing constructed by the convolution of $P^1_{e^+}$ and $P_\gamma$:

$$P_{CS}(i) = \sum_{j=0}^{i} P_\gamma(j) P_{e^+}^1(i-j) \tag{A2}$$

The first two terms in Eq. (A1) correspond to β+ decay and the third term to EC.

4. Lastly, the resulting convoluted histogram was Gaussian broadened [32, 33] to obtain a realistic representation of the spread in scintillation photon yield per keV γ-ray energy absorbed by the NaI scintillator.

In Fig. 9 the β+ and EC contributions are shown separately, including their respective Compton continua. In addition to the peaks at 511, 1022, and 1275 keV also peaks at the true coincident energies 1786 and 2297 keV are visible. The intensities of these peaks reflect the probabilities that one, two, or all three γ-ray photons of a single $^{22}$Na decay are detected by our setup.

In the comparison between simulations and measurements in Fig. 4, one notices that there is some mismatch between the sets of five peak positions. We interpret this difference being related to light collection properties of the well-counter. With the source positioned at the bottom of the well, the photons are emitted isotropically from the center of the detector. The mean free path of the annihilation photons is smaller than that of the deexcitation photons. If the efficiency of light collection at the photo-cathode has some dependency on the position in the crystal where it scintillates, then the combination of these effects will show up as peak shifts that are dependent on the energy of the photon(s) involved.

TABLE I. Calculation of the total number of antineutrinos produced per fission at the start and the end of the burn-up cycle. These totals (in bold) are sums weighted by fissioning nuclides fractions that were derived from known fission neutron fractions.

| Nuclide | Fission neutron yield | Fission antineutrino yield | Fission neutron fraction at start | Fission neutron fraction at end | Fissioning nuclide fraction at start | Fissioning nuclide fraction at end | Antineutrino yield × nuclide fraction at start | Antineutrino yield × nuclide fraction at end |
|---|---|---|---|---|---|---|---|---|
| $^{235}$U | 2.5 | 5.58 | 67% | 42% | 65% | 41% | 3.61 | 2.27 |
| $^{238}$U | 1.7 | 6.69 | 8% | 8% | 11% | 11% | 0.76 | 0.76 |
| $^{239}$Pu | 2.9 | 5.09 | 25% | 50% | 21% | 42% | 1.06 | 2.13 |
| $^{241}$Pu | | 5.89 | <1% | <1% | 3% | 6% | 0.18 | 0.36 |
| sum | | | 100% by definition | | | | **5.61** | **5.52** |

TABLE II. Reactor-status periods for two measurement series with the $^{22}$Na source.

| Series | Reactor status | Period (DOY 2013) | From | Until |
|---|---|---|---|---|
| 1 | ON1 | 3 - 43 | 3-Jan-2013 | 12-Feb-2013 |
| 1 | OFF1 | 58 - 110 | 27-Feb-2013 | 20-Apr-2013 |
| 1 | ON2 | 117 - 140 | 27-Apr-2013 | 20-May-2013 |
| 2 | ON3 | 283 - 314 | 10-Oct-2013 | 10-Nov-2013 |
| 2 | OFF2 | 320 - 361 | 16-Nov-2013 | 27-Dec-2013 |
| 2 | ON4 | 367 - 405 | 2-Jan-2014 | 9-Feb-2014 |

TABLE III. Regression results for the parametrization of Eq. (2) of the two measurement series with the $^{22}$Na source. Bracketed numbers behind fitted parameter values denote the 1-sigma Poisson counting error in the least significant digit shown. Parameter $N$ is the number of fit parameters. The last column shows p-values for the regressions on the same data but ignoring reactor-status dependency.

| Series | RoI | Eq. (2a) aver | Eq. (2b) $dip_{ampl}$ | Eq. (2b) $dip_{decay}$ ($day^{-1}$) | Eq. (2c) frac | Eq. (2d) $trend_{rate}$ ($10^{-6}\ day^{-1}$) | $N$ | $\chi^2$ | dof | Eq. (2c) $P$ | Eq. (2e) $P$ |
|---|---|---|---|---|---|---|---|---|---|---|---|
| 1 | TOT | 1.00038 (2) | −0.00086 (3) | 0.140 (13) | −0.000300 (17) | −2.5 (2) | 5 | 155 | 113 | 0.0052 | 2×10$^{-45}$ |
| 1 | MED | 1.00013 (4) | | | 0.00020 (5) | −13.0 (6) | 3 | 146 | 115 | 0.027 | 0.0021 |
| 1 | HI | 1.00035 (2) | | | −0.00026 (3) | −0.5 (4) | 3 | 173 | 115 | 4×10$^{-4}$ | 8×10$^{-12}$ |
| 2 | TOT | 1.00057 (14) | −0.00071 (12) | 0.046 (16) | −0.000305 (19) | −1.1 (3) | 5 | 134 | 108 | 0.045 | 4×10$^{-34}$ |
| 2 | MED | 1.00032 (4) | | | 0.00007 (6) | −22.9 (7) | 3 | 153 | 110 | 0.0040 | 0.0038 |
| 2 | HI | 1.00056 (3) | | | −0.00028 (3) | 4.7 (4) | 3 | 136 | 110 | 0.046 | 1×10$^{-7}$ |

TABLE IV. Fractional countrate change values for the three Regions of Interest, for the two measurement series and their weighted averages.

| Series | $\left(\frac{\Delta A}{A}\right)_{TOT} \times 10^4$ | $\left(\frac{\Delta A}{A}\right)_{MED} \times 10^4$ | $\left(\frac{\Delta A}{A}\right)_{HI} \times 10^4$ |
|---|---|---|---|
| 1 | −3.00 ± 0.20(stat) | +2.0 ± 0.6(stat) | −2.6 ± 0.4(stat) |
| 2 | −3.05 ± 0.21(stat) | +0.7 ± 0.7(stat) | −2.8 ± 0.4(stat) |
| 1 & 2 | −3.02 ± 0.14(stat) ± 0.07(syst) | +1.44 ± 0.42(stat) ± 0.07(syst) | −2.70 ± 0.26(stat) ± 0.04(syst) |

FIG. 1. Schematic cross section of the relevant parts of the reactor building at Koeberg. Total thickness of concrete layers between reactor vessel and seismic vault is approx. 8 m.

FIG. 2. Difference spectrum between background during reactor ON and background during reactor OFF (dots). For comparison, also shown is the background during reactor OFF divided by a factor 20 (line).

FIG. 3. Background countrate in the energy range 320–2452 keV with reactor OFF. The X-axis represents time expressed in day-of-year in the year 2015.

FIG. 4. Gamma-ray spectrum of a $^{22}$Na source inside our well counter, measured and simulated by MCNPX. The insert shows the $^{22}$Na decay scheme.

FIG. 5. Daily averaged countrates in RoI TOT for measurement series 1 (a) and 2 (b). The normalization to relative countrate was done with the average countrate over that period. The various states of the reactor are shown above the X-axis. Label 'T' stands for Transition between ON and OFF states. During period 'BG' in series 1 the source was removed to take some background data. The error bars represent the 1-sigma Poisson counting error. The solid curve is obtained from a functional description of the dip structure in a fit to the countrate data.

FIG. 6. Relative countrates in RoIs TOT, MED, and HI for measurement series 1, in panels (a), (b), and (c), respectively, and for measurement series 2, in panels (d), (e), and (f), respectively. These six panels show all the data collected with the $^{22}$Na source, after correction for the dip structure and linear trends to best visualize the reactor-status effect. Solid horizontal lines represent the average value during reactor-status ON and OFF periods.

FIG. 7. Countrate of a $^{60}$Co source in the RoI $E_\gamma = 2376–2719\,keV$, during a reactor-status change from ON, through a transition T, to OFF.

FIG. 8. Decay scheme of $^{50}$V to $^{50}$Ti. The transition to $^{50}$Ti(1554) via $\beta^+$ decay is energetically allowed if the captured electron antineutrino has an incoming energy $E > 368\,keV$.

FIG. 9. MCNPX simulation of the combined response of the NaI well-geometry crystal in our setup to the photoelectric effect and Compton scattering of γ-rays from the decay of $^{22}$Na, for EC and $\beta^+$ decay separately. RoI MED, centered around the $E_\gamma = 1275\,keV$ deexcitation peak, captures γ-rays from these decay modes with probabilities in this ratio: $EC:\beta^+ \approx 1:3$.

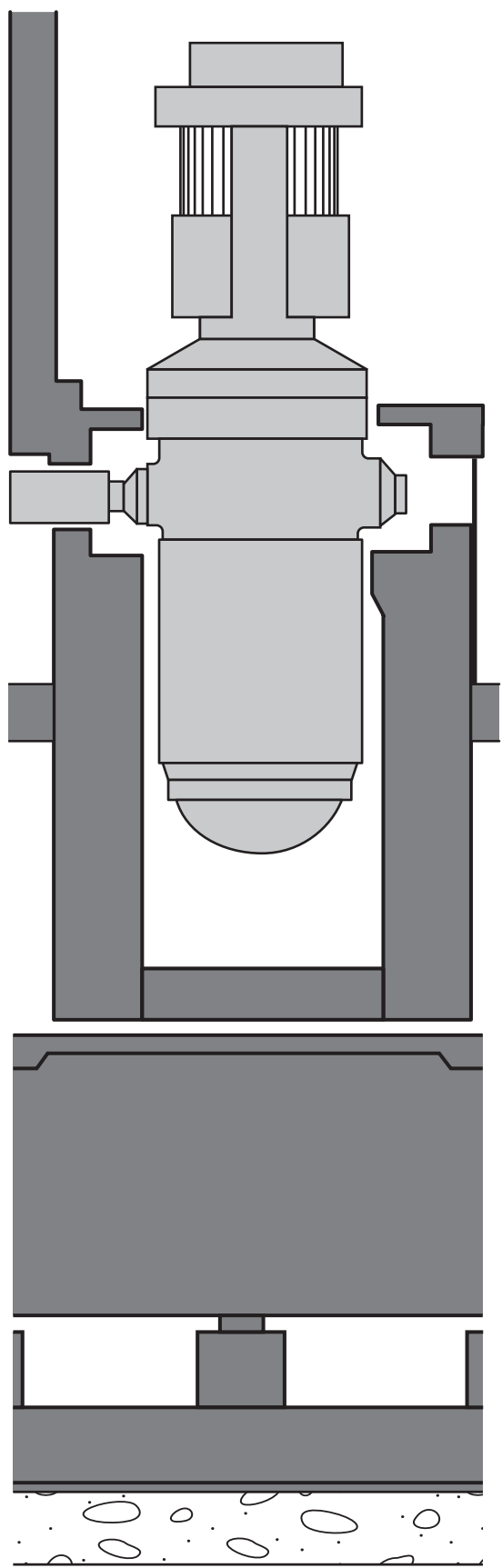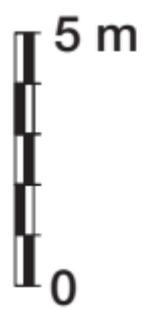

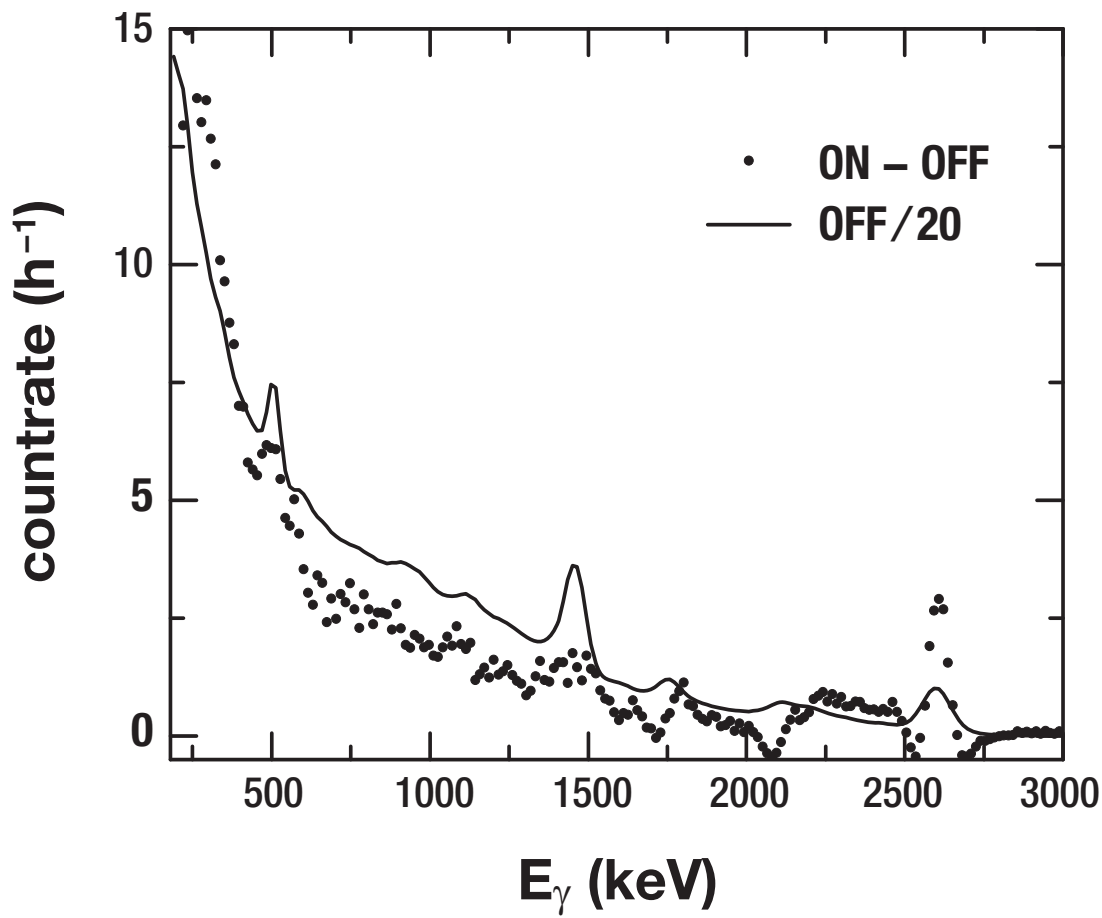

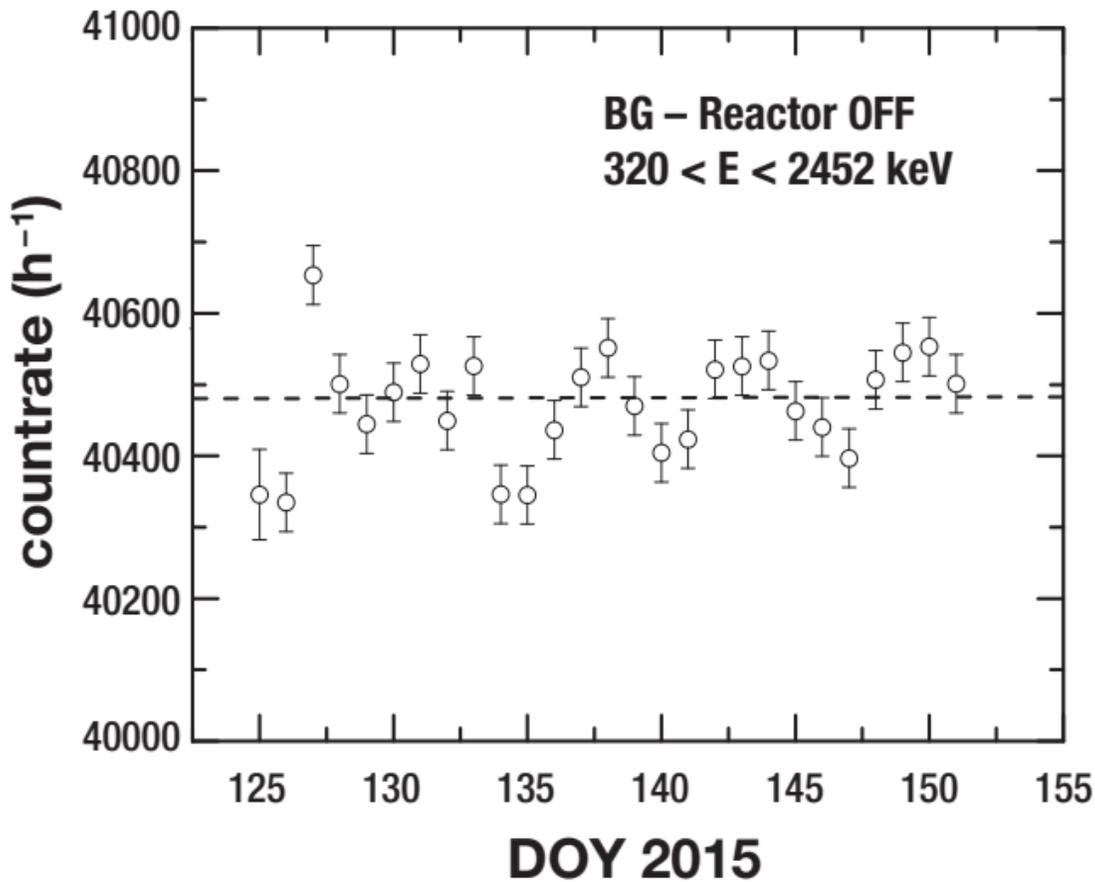

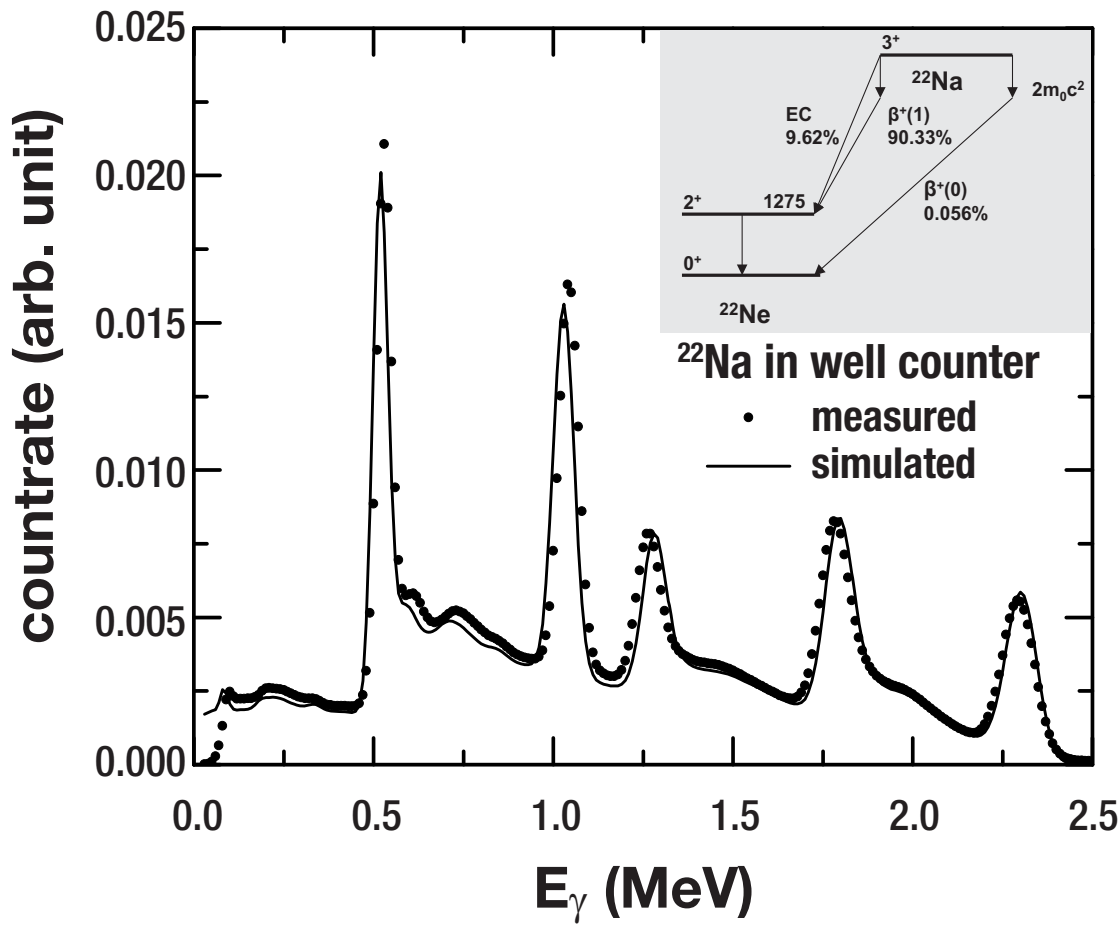

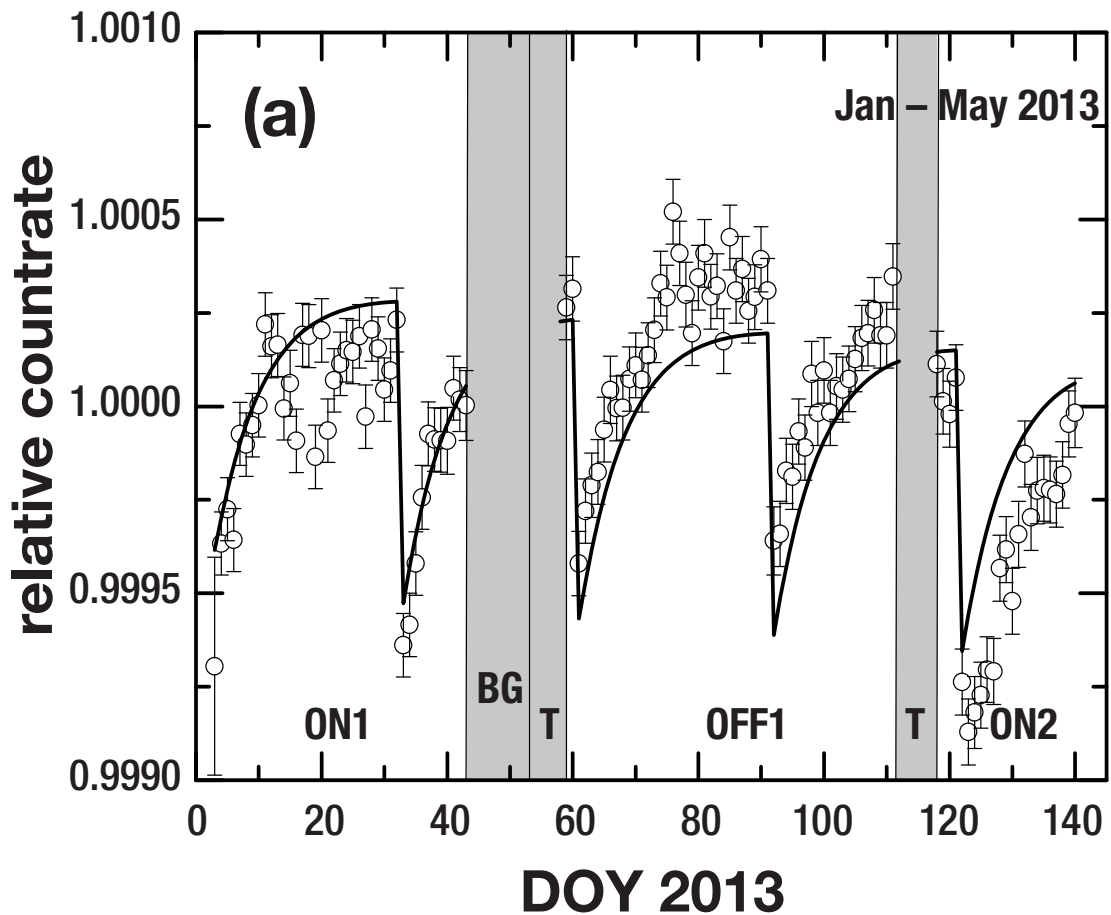

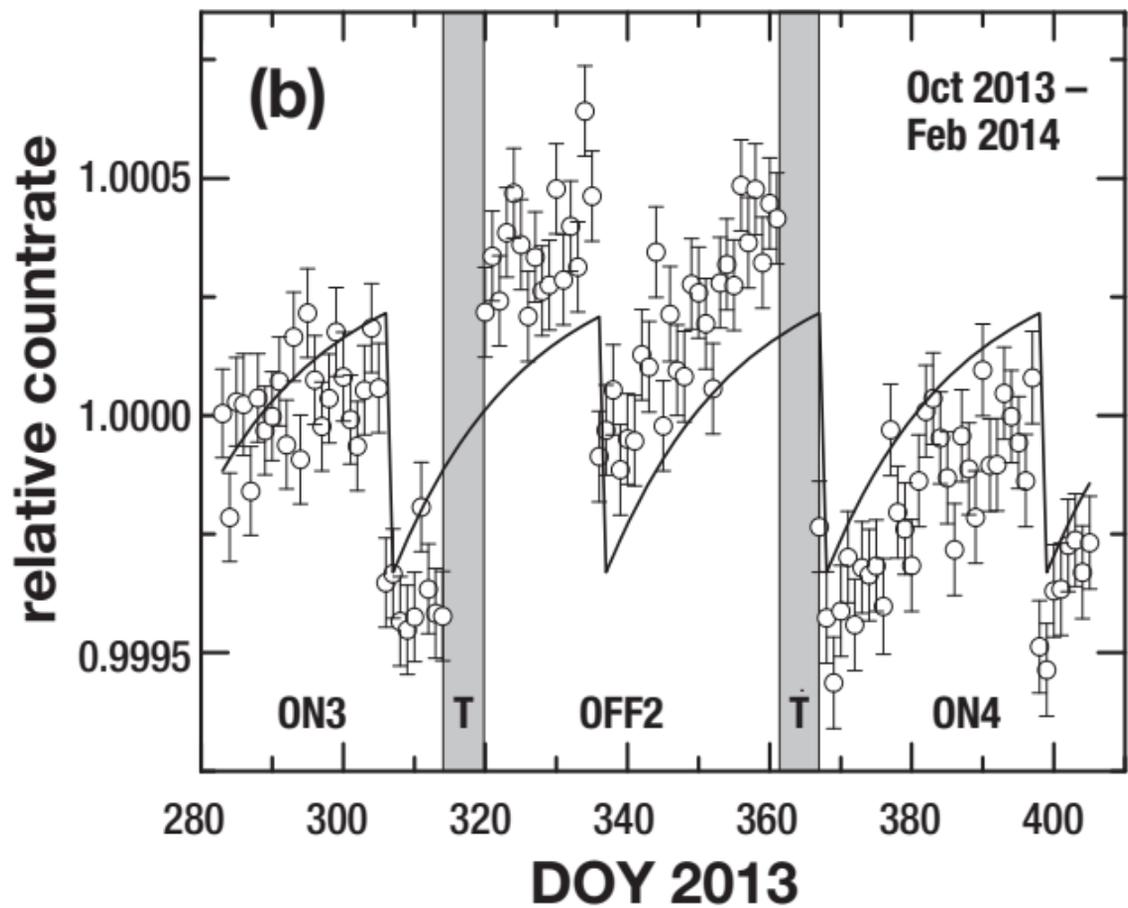

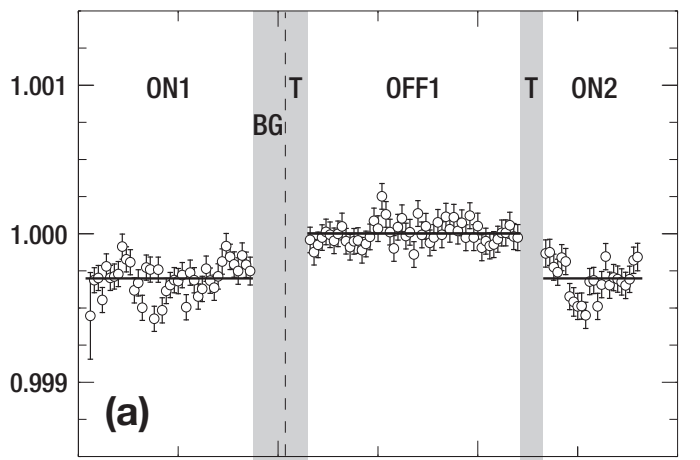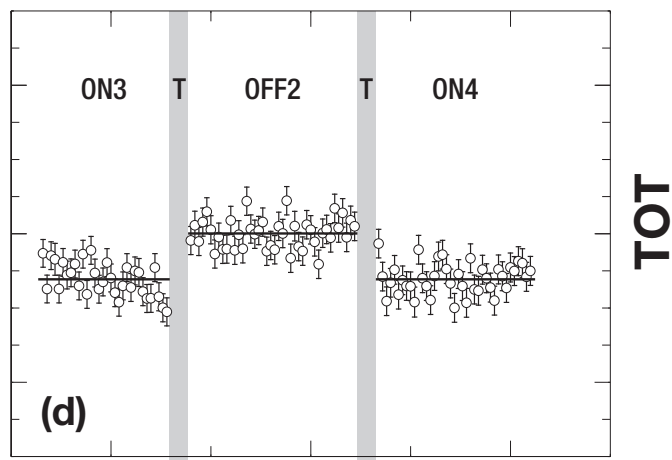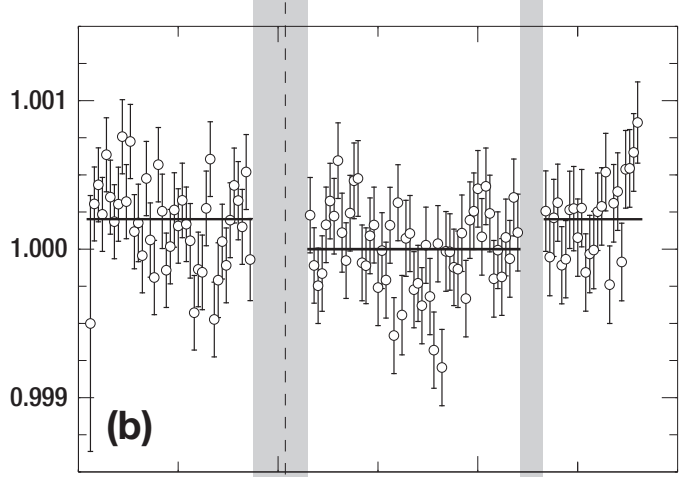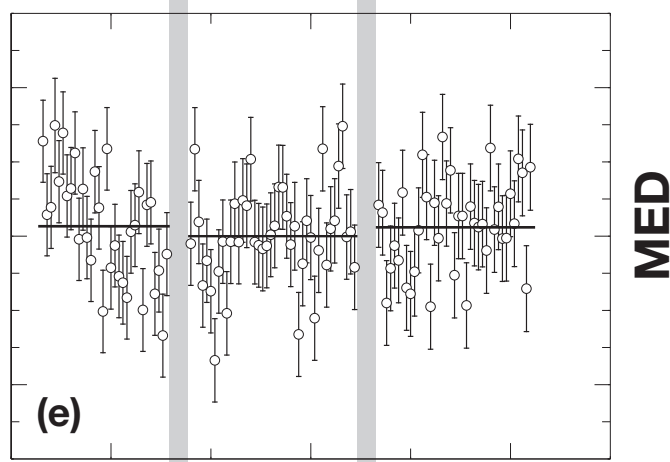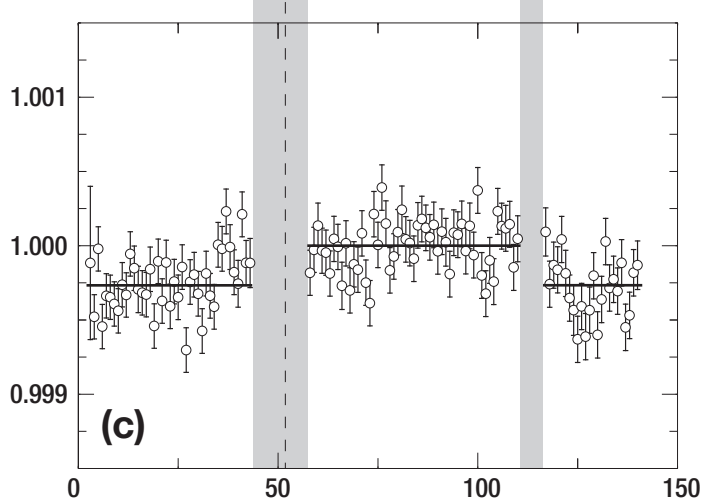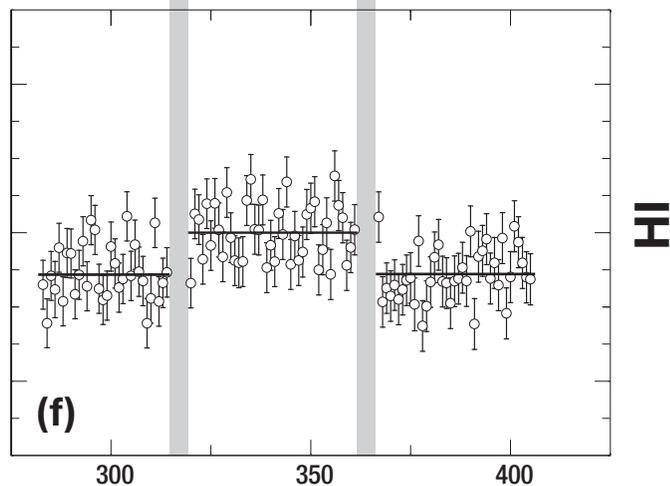

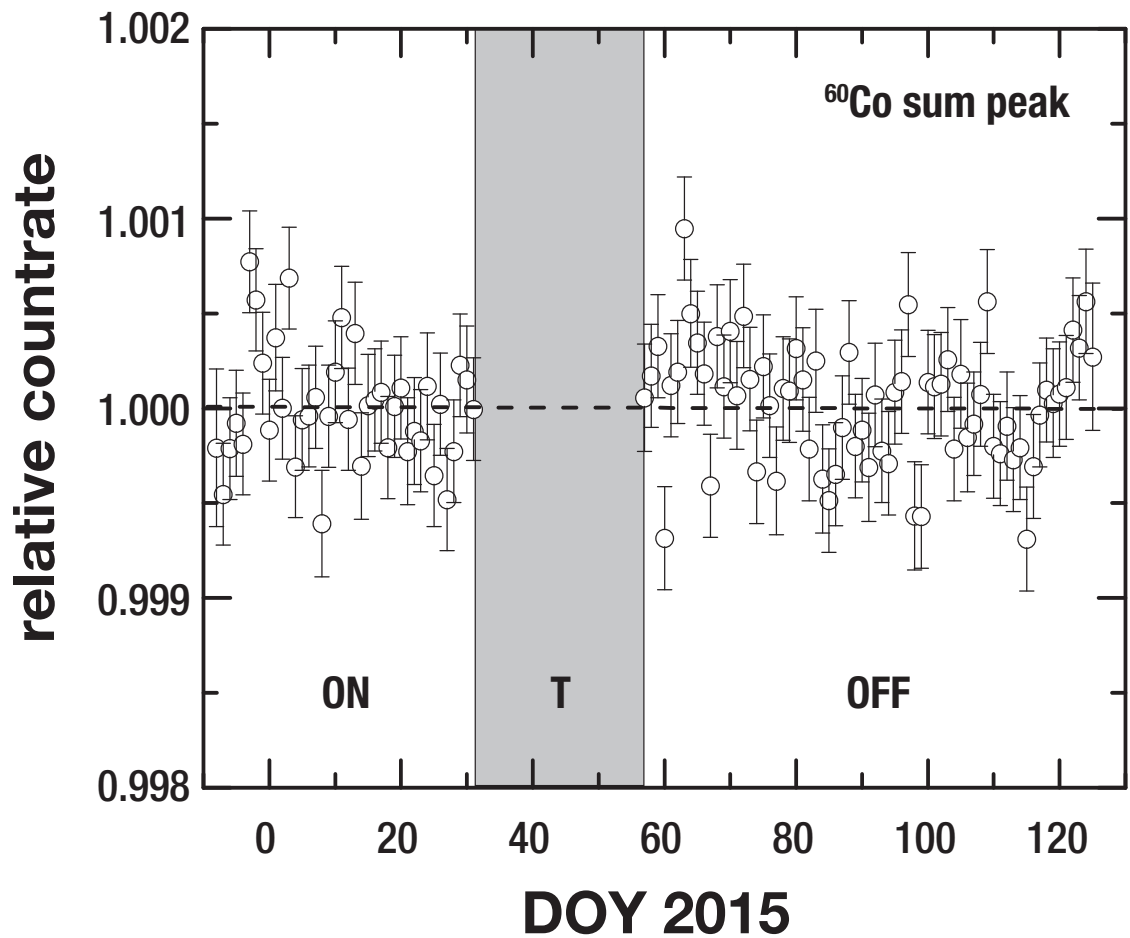

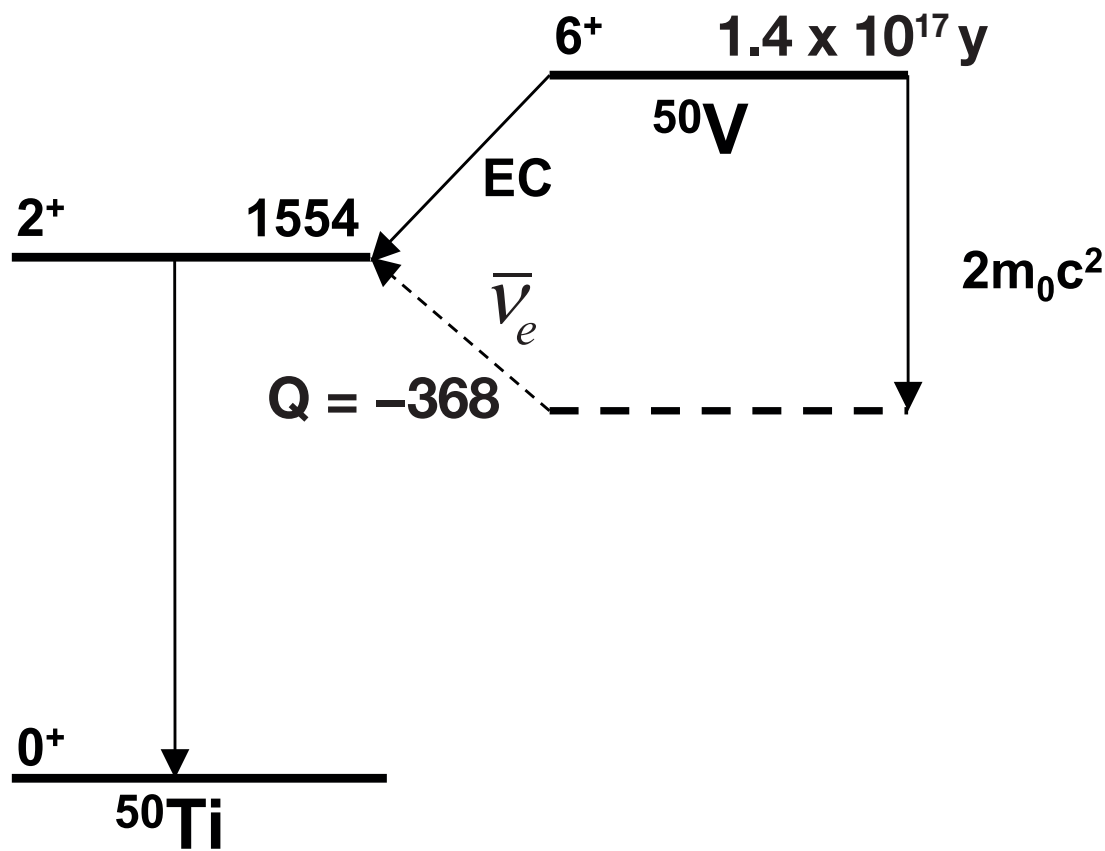

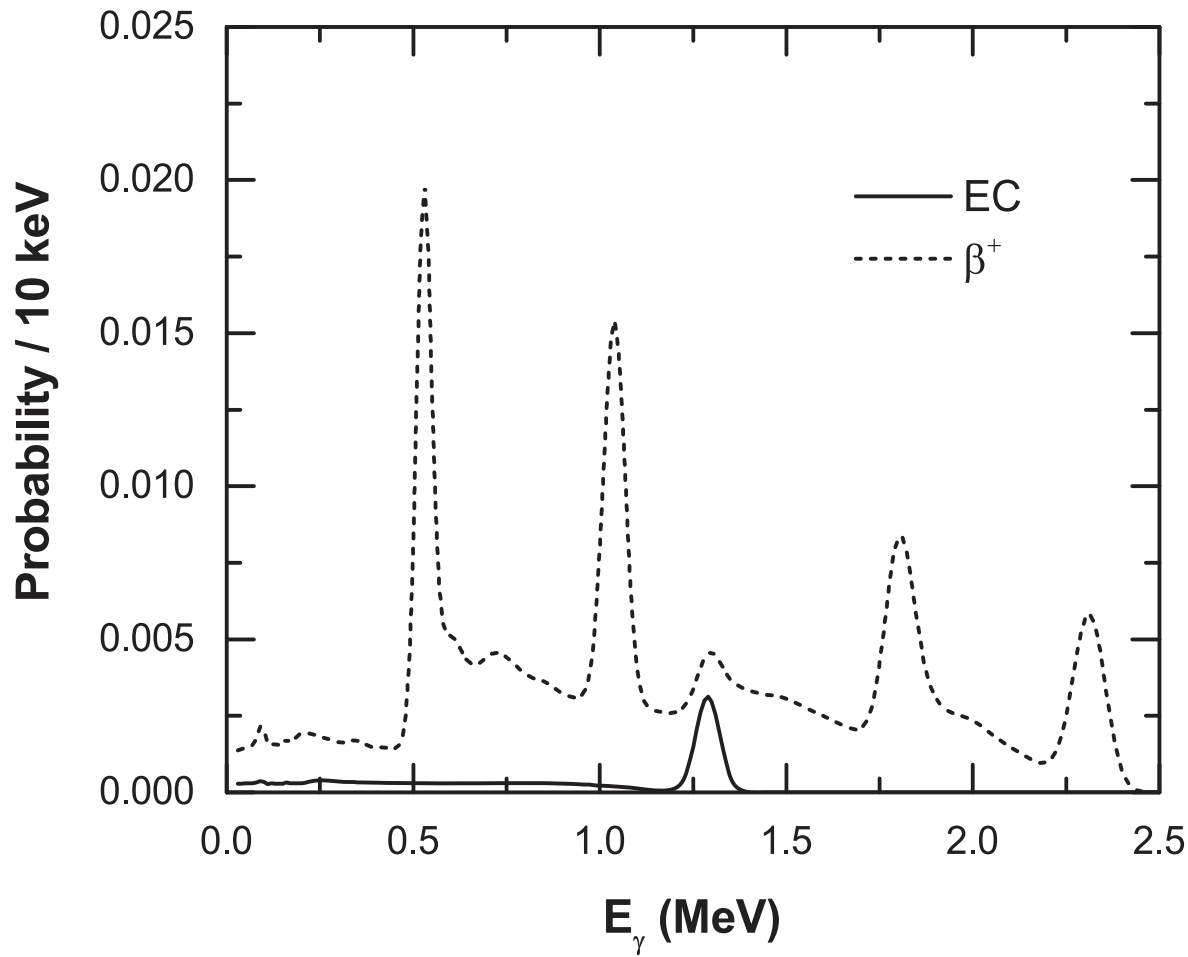